\documentclass[aps,nofootinbib,preprint,tightenlines]{revtex4}

\usepackage{graphicx}
\usepackage{amsmath}
\usepackage{amssymb}
\usepackage{bm} 

\newcommand{\sing}{^1{\rm S}_0}
\newcommand{\trip}{^3{\rm S}_1}

\newcommand{\mpic}{M_\pi^{crit}}
\newcommand{\st}{{s,t}}

\renewcommand{\i}{\mathrm{i}}
\newcommand{\e}{\mathrm{e}}
\newcommand{\dd}{\mathrm{d}}

\newcommand{\halb}{\frac{1}{2}}
\newcommand{\lp}{{\ell^{\prime}}}

\newcommand{\oneover}[1]{\frac{1}{#1}}
\newcommand{\Zdrei}{\mathbb{Z}^3}
\newcommand{\Ct}{\tilde{C}}

\newcommand{\mylj}[1]{\ell(#1)}
\newcommand{\lj}{\mylj{j}}
\newcommand{\lJ}{\mylj{J}}
\newcommand{\gsim}{\hspace*{0.2em}\raisebox{0.5ex}{$>$}
     \hspace{-0.8em}\raisebox{-0.3em}{$\sim$}\hspace*{0.2em}}

\def\mqo2{{\!\!\!}}

\begin{document}


\preprint{HISKP-TH-10-23,\
INT-PUB-10-039}
\title{The triton in a finite volume}
\author{Simon Kreuzer}

\author{H.-W. Hammer}
\affiliation{Helmholtz-Institut f\"ur Strahlen- und Kernphysik (Theorie)\\
and Bethe Center for Theoretical Physics,
 Universit\"at Bonn, 53115 Bonn, Germany\\}

\date{\today}

\begin{abstract}
  Understanding the volume dependence of the triton binding energy is an 
  important step towards lattice simulations of light nuclei. 
  We calculate the triton binding energy in a
  finite cubic box with periodic boundary conditions 
  to leading order in the pionless effective
  field theory. Higher order corrections are estimated and the 
  proper renormalization of our results is verified explicitly.
  We present results for the physical triton as well as for
  the pion-mass dependence of the triton spectrum near the ``critical''
  pion mass, $\mpic \approx 197$~MeV, where chiral effective field theory 
  suggests that the nucleon-nucleon scattering lengths in the $\sing$- and 
  $\trip$-channels diverge simultaneously.
  An extension of the L\"uscher formula to the three-body system is
  implicit in our results.
\end{abstract}

\maketitle

Although Quantum Chromodynamics (QCD) is widely accepted as the
underlying theory of strong interactions, ab initio calculations of
nuclear properties in Lattice QCD remain one of the largest
theoretical challenges in the Standard Model
\cite{Beane:2008dv,Beane:2010em}.  In nuclear physics, the relevant
degrees of freedom are pions and nucleons. Traditionally, their
interactions are described via phenomenological potentials fitted to
the nucleon-nucleon scattering data. More recently, the advent of
model independent Effective Field Theory (EFT) approaches has allowed
for accurate calculations of low-energy nuclear physics observables
with a direct connection to QCD via its symmetries
\cite{Beane:2000fx,Bedaque:2002mn,Epelbaum:2005pn,Epelbaum:2008ga}.

In Lattice QCD, the QCD path integral is evaluated in a discretized
Euclidean space-time using Monte Carlo
simulations~\cite{Wilson:2004de}.  However, this approach requires a
large numerical effort that strongly constrains the parameters of the
simulation. In particular, one is at present forced to use relatively
small finite volumes. The energy of states calculated in the finite
volume is shifted relative to the infinite volume limit
\cite{Luscher:1985dn}.  This shift has to be taken into account when
extracting physical observables from lattice simulations.  The finite
volume also allows for the extraction of scattering observables away
from kinematic thresholds.  By generalizing a quantum mechanical
result to field theory, L{\"u}scher showed that the volume dependence
of two-body energy states encodes the infinite volume scattering phase
shift~\cite{Luscher:1990ux} as well as resonance
properties~\cite{Luscher:1991cf}. An extension of these results to the
three-body sector is required for the simulation of light nuclei and
their scattering properties in Lattice QCD.

This necessity
motivates our investigation of the volume dependence of the triton 
binding energy within the pionless EFT. This EFT is valid for
processes with typical momenta below the pion mass and 
has succesfully been applied to describe the properties of light nuclei
\cite{Beane:2000fx,Bedaque:2002mn,Epelbaum:2008ga}.
To leading order, the triton properties are determined by the 
nucleon-nucleon scattering lengths in the $\sing$ and $\trip$ channels
and a Wigner SU(4) symmetric three-body force \cite{Bedaque:1999ve}.
The triton has been considered in pionless EFT in a nuclear lattice 
formalism but the volume dependence was not 
investigated \cite{Borasoy:2005yc}.
Higher-order corrections to the amplitude including the ones due to 2N
effective range terms can be treated perturbatively 
\cite{Efimov:1991aa}.  The inclusion of such 
corrections in the pionless EFT has been studied extensively 
\cite{Hammer:2000nf,Bedaque:2002yg,Griesshammer:2004pe,Platter:2006ad}.
Our strategy to calculate the modification of the triton
in a cubic box has previously been employed to investigate bound states
of three bosons with large scattering length
\cite{Kreuzer:2008bi,Kreuzer:2009jp}.  In a different
approach, Epelbaum and collaborators have calculated the energy of the
triton in a finite volume by implementing a discretized version of
chiral EFT on a lattice~\cite{Epelbaum:2009zsa}.
The correlation function for the three-nucleon system in the triton channel 
has also been calculated in Lattice QCD recently~\cite{Beane:2009gs}, 
but because of the relatively large uncertainties 
no triton properties could be extracted. 

Another motivation for our work is the observation that QCD lies close
to an infrared renormalization group limit cycle~\cite{Braaten:2003eu}.  
The pion mass dependence of the nucleon-nucleon scattering lengths
from chiral EFT is compatible with simultaneously diverging
nucleon-nucleon scattering lengths in the $\sing$ as well as in the
$\trip$ channel near a critical pion mass $M_\pi^{crit} \approx
197$~MeV~\cite{Beane:2001bc,Beane:2002xf,Epelbaum:2002gb}.  As a
consequence, it was conjectured that QCD could be tuned to the
limit cycle by slightly changing the up and down quark masses
~\cite{Braaten:2003eu}.  In this scenario, 
excited states of the triton would appear near the critical
quark masses. At the critical point itself, the triton would have
infinitely many excited states with an accumulation point at
threshold. This is a signature of the universal Efimov effect 
which appears in three-body systems with resonant short-range
interactions \cite{Efimov-70}. It has recently
been observed experimentally in atomic
systems~\cite{Ferlaino:2010} and many nuclear systems can be
described in an expansion around the ideal Efimov limit
\cite{Braaten:2004rn,Hammer:2010kp}.
The large scattering lengths make the region near
the critical pion mass accessible to the pionless EFT,
whose input parameters for various pion masses can
be obtained from chiral EFT.  This program has successfully
been carried out to next-to-next-to-leading order
(N$^2$LO) in the infinite
volume~\cite{Epelbaum:2006jc,Hammer:2007kq}.  
When investigating this pion mass region
with Lattice QCD, it is again desirable to have control over the
effects of the finite volume. In particular, the influence of the
finite volume on the infrared limit cycle is not known.

To leading order in the pionless EFT,
the Lagrangian for a system of three nucleons
can be written as \cite{Bedaque:1999ve}
\begin{equation}\begin{split}
  \label{eq:lagrangian}
  \mathcal{L} =& \,N^\dagger\left(i\partial_t + \halb\nabla^2\right)N
    +\frac{g_t}{2}\,{t_j}^\dagger {t_j}
    +\frac{g_s}{2}\,{s_A}^\dagger {s_A} \\
    &-\frac{g_t}{2}\,\left[ {t_j}^\dagger \,(N^T\tau_2\sigma_j\sigma_2 N) 
     + \text{h.c.}\right]
     -\frac{g_s}{2}\,\left[{s_A}^\dagger \, (N^T \sigma_2 \tau_A \tau_2 N) 
     + \text{h.c.}\right] 
    + \mathcal{L}_3\,,
\end{split}\end{equation}
where the units have been set such that $\hbar=m=1$ and $\sigma_j$
($\tau_A$) are the Pauli matrices acting in spin (isospin) space. The degrees
of freedom in this Lagrangian are the nucleon field~$N$ and two
auxiliary dinucleon fields, $t_j$ and $s_A$. The field~$t_j$ ($s_A$)
corresponds to two nucleons in the $\trip$ ($\sing$)-channel. The
SU(4)-invariant three-body interaction contained in~$\mathcal{L}_3$ is 
proportional to $(N^\dagger N)^3$.
It can be written as a dinucleon-nucleon contact interaction with a 
dimensionless coupling constant~$H(\Lambda)$~\cite{Bedaque:1999ve},
\begin{equation}\begin{split}
\label{3bod}
{\cal L}_3 =& -\frac{2H(\Lambda)}{\Lambda^2}
  \bigg( g_t^2 N^\dagger 
(t_j\sigma_j)^\dagger (t_i \sigma_i) N 
 +\frac{g_t g_s}{3} \left[ N^\dagger (t_j\sigma_j)^\dagger
(s_A\tau_A) N + \text{h.c.} \right] \\
& \qquad\qquad\quad + g_s^2 N^\dagger (s_A\tau_A)^\dagger (s_B\tau_B) N \bigg)\,,
\end{split}\end{equation}
where~$\Lambda$ is the momentum cutoff.
Finite-range corrections can be incorporated as
higher orders in the EFT 
\cite{Hammer:2000nf,Bedaque:2002yg,Griesshammer:2004pe,Platter:2006ad}
but will not be considered here. 
The coupling constants~$g_\st$ can be matched to the two-body
scattering lengths~$a_\st$ in the corresponding channel or, in the
case of $g_t$, to the deuteron binding energy. The three-body
coupling~$H(\Lambda)$ approaches an ultraviolet limit cycle for large
$\Lambda$. The analytic dependence of $H$ on the
cutoff can be expressed as
\begin{equation}\label{eq_fw_H}
H(\Lambda)=\frac{\cos\left[s_0\log(\Lambda/\Lambda_\ast)+\arctan s_0\right]} 
{\cos\left[s_0\log(\Lambda/\Lambda_\ast)-\arctan s_0\right]}\,,
\end{equation}
where $s_0=1.00624...$ and the phase~$\Lambda_\ast$ has to be 
fixed from a three-body datum. The low-energy constant 
$\Lambda_\ast$ is also known as the
\lq\lq three-body parameter''. This renormalization procedure 
remedies the incorrect ultraviolet behavior of the EFT. Once $\Lambda_\ast$ has
been fixed from a three-body datum, all other low-energy three-body
observables can be predicted. For more details on the infinite
volume case we refer the reader to the literature
\cite{Bedaque:1999ve,Hammer:2000nf,Bedaque:2002yg,Griesshammer:2004pe,
Platter:2006ad}.

We now consider placing the system inside a cubic box.
The finite volume modifies the
infrared regime due to momentum quantization. In a finite cubic volume
with edge length~$L$ and periodic boundary conditions, the allowed
momenta are given by~$\frac{2\pi}{L}\vec{n},\,\vec{n}\in\Zdrei$,
yielding a low-momentum scale~$\frac{2\pi}{L}$ corresponding to the
minimal accessible non-zero momentum.  As long as this scale is small
compared to the high-energy scale given by the momentum
cutoff~$\Lambda$, the ultraviolet behavior of the amplitudes is
unaffected by the finite volume and the renormalized values for the
coupling constants obtained in the infinite volume can be used for
finite volume calculations. We will explicitly verify that this claim
holds for our results.

\begin{figure}[t]
  \centering
  \includegraphics[width=.98\linewidth]{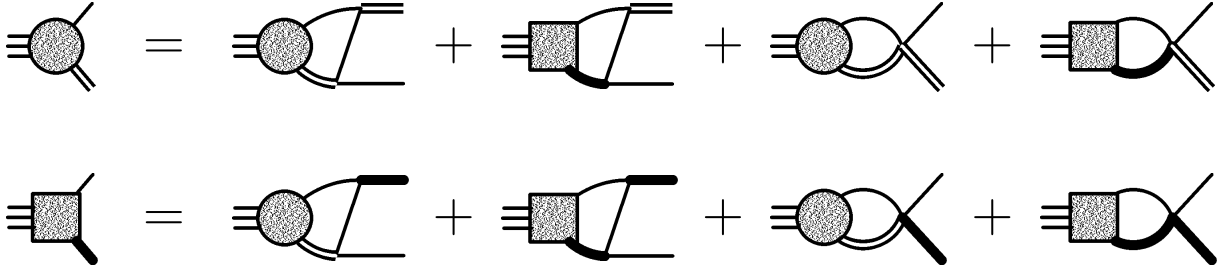}
  \caption{Integral equation for the triton
    amplitudes $\mathcal{F}_t$ (shaded circle) and $\mathcal{F}_s$
    (shaded square). Single lines
    denote nucleons, double lines indicate
    dinucleons in the $\trip$ channel and thick solid lines 
    denote dinucleons in the $\sing$ channel.} 
  \label{fig:inteq}
\end{figure}
The properties of the triton are determined by the bound state
amplitude in the spin-$1/2$ channel which is given by the
integral equation in Fig.~\ref{fig:inteq}.
This equation has non-trivial solutions only for negative 
energies $E_3$ whose absolute values correspond to the
binding energies.
The triton amplitude has two components $\mathcal{F}_t$ (shaded circle) 
and $\mathcal{F}_s$ (shaded square) that correspond to the outgoing
dinucleon being in the $\trip$ and $\sing$ channels.
The single lines denote nucleons, while the double and thick solid 
lines indicate full dinucleon propagators in the $\trip$ and $\sing$ channels,
respectively.
The amplitude gets contributions from one-nucleon exchange as well
as from the three-body interaction in ${\cal L}_3$. The loop momenta in this
equation are quantized in a finite volume as described above. 
The full propagator for the dinucleon fields is obtained by dressing the 
bare propagator from Eq.~\eqref{eq:lagrangian} with nucleon loops
to all orders. 
This yields an geometric series of diagrams that can be evaluated 
analytically:
\begin{equation}
  \label{eq:deutprop}
  D_\st(p_0,\vec{p}) = \frac{8\pi}{g_\st^2}\bigg[-\oneover{a_\st}
   +\sqrt{-p_0+\vec{p}^{\,2}/4-i\epsilon}- 
    \sum_{\substack{\vec{\jmath}\in\Zdrei \\ \vec{\jmath}\neq\vec{0}}}
    \oneover{|\vec{\jmath}|L} \e^{-|\vec{\jmath}|L\sqrt{-p_0+\vec{p}^{\,2}/4
    -i\epsilon}}
   \bigg]^{-1}.
\end{equation}
The result reproduces the infinite volume dinucleon propagator except
for a volume dependent term that vanishes in the limit $L\to\infty$.

Using the Feynman rules from the Lagrangian in
Eqs.~\eqref{eq:lagrangian} and \eqref{3bod} 
and the full dimer propagators from above, we
can explicitly write down the coupled integral equations for the
triton amplitude from Fig.~\ref{fig:inteq}. It
involves integrations over the loop energy and sums over the quantized
loop momenta. The integrations are performed by virtue of the residue
theorem while the sums over the quantized momenta are rewritten into
sums of integrals using Poisson's resummation formula
$\sum_{\vec{m}\in\Zdrei} \delta^{(3)}(\vec{x}-\vec{m}) =
\sum_{\vec{n}\in\Zdrei}\e^{2\pi\i\,\vec{x}\cdot\vec{n}}$, which is
understood to be used under an integral. After projection on
the quantum numbers of the triton,  we have
\begin{widetext}\begin{equation}\label{eq:inthom}
\begin{pmatrix}\mathcal{F}_t(\vec{p}) \\ \mathcal{F}_s(\vec{p})\end{pmatrix} =
 \oneover{\pi^2} \sum_{\vec{n} \in \Zdrei} \int_0^\Lambda \dd^3 y\,
 \e^{\i L \vec{n}\cdot\vec{y}} \left[ \mathbf{M}_2(\vec{y})\,
   \mathcal{Z}(\vec{p}, \vec{y}) + \mathbf{M}_3(\vec{y})\,
   \frac{2H(\Lambda)}{\Lambda^2}\right]
 \begin{pmatrix} \mathcal{F}_t(\vec{y}) \\
    \mathcal{F}_s(\vec{y}) \end{pmatrix}
\end{equation}\end{widetext}
where $\mathcal{Z}(\vec{p}, \vec{y}) =
\bigl[p^2+\vec{p}\cdot\vec{y}+y^2-E_3\bigr]^{-1}$. 
The matrix-valued functions $\mathbf{M}_2(\vec{y})$ and 
$\mathbf{M}_3(\vec{y})$ are given by
\begin{widetext}\begin{equation}\label{eq:M23}
\mathbf{M}_2(\vec{y}) = \begin{pmatrix} -d_t(\vec{y}) & 3d_s(\vec{y}) \\ 
3d_t(\vec{y}) & -d_s(\vec{y})\end{pmatrix}\,,
\qquad
\mathbf{M}_3(\vec{y}) =  \begin{pmatrix} -d_t(\vec{y}) & d_s(\vec{y}) \\ 
d_t(\vec{y}) & -d_s(\vec{y})\end{pmatrix}\,,
\end{equation}\end{widetext}
where $d_\st(\vec{y})=
(g_\st^2/8\pi)D_\st(E_3-\vec{y}^{\,2}/2,\vec{y})$ depends only 
on the absolute value of~$\vec{y}$. 
The integral equations~\eqref{eq:inthom} 
obey Wigner SU(4) symmetry in the ultraviolet. 
As a consequence, a SU(4)-symmetric three-body interaction 
with running coupling $H(\Lambda)$ is sufficient for
renormalization~\cite{Bedaque:1999ve}.

In a finite cubic box, spherical symmetry is broken down to cubic
symmetry. As a consequence, the infinitely many irreducible
representations of the double cover of the rotational group, SU(2),
become reducible in terms of the eight irreducible representations of
the double cover of the cubic group, $^2O$. The triton has $j=1/2$.
This partial
wave is contained in the $G_1^+$ representation, which also contains
$j=7/2, 9/2, \dots$~\cite{Johnson:1982yq}. Assuming that the triton
amplitude in finite volume transforms under the $G_1^+$
representation, it is possible to decompose it into the different
partial waves~\cite{Bethe47,Bernard:2008ax} as
\begin{equation}
  \label{eq:kubic}
  \begin{split}
    \mathcal{F}(\vec{y}) &= \sum_{j=\halb,\frac{7}{2},\dots}^{(G_1^+)} \sum_t 
       F^{(j,t)}(y) \sum_{m_j} \Ct_{jtm_j} |jm_j\rangle.
  \end{split}
\end{equation}
The coefficients $\Ct_{jtm_j}$ can be determined by explicitly
decomposing the reducible representations~\cite{Bernard:2008ax} and
the sum over $t$ is needed if a partial wave is contained more than
once. Since this is not the case for partial waves less than~$13/2$,
we will omit this index in the following. The vectors $|jm_j\rangle$
are given by $|jm_j\rangle = \sum_{m,s} C^{jm_j}_{\lj m \halb s}
|\lj m\rangle\otimes|\halb s\rangle$, where the $C$'s are
Clebsch-Gordan coefficients, $|\ell m\rangle$ is a spherical harmonic
and $|\halb s\rangle$ is a spin-$1/2$ spinor. The sign in
$\lj=j\pm\halb$ has to be chosen such that $\lj$ is even in order
to get the positive parity of the triton.

Projecting out the $J$th partial wave and performing the angular
integrations in Eq.~\eqref{eq:inthom} yields an infinite set of
coupled equations:
\begin{widetext}
\begin{equation}\begin{split}
  \label{eq:partwaves}
  \begin{pmatrix}F^{(J)}_t(y) \\ F^{(J)}_s(y)\end{pmatrix} &= 
    \frac{4}{\pi} \int_0^\Lambda \frac{\dd y\, y^2}{2\lJ+1} 
    \sum_j^{(G_1^+)}\left[\mathbf{M}_2(y)\,
     Z^{(\lJ)}(p,y) + \mathbf{M}_3(y)\,
      \frac{2H(\Lambda)}{\Lambda^2}\,\delta_{\lJ,0}
    \right]
    \begin{pmatrix}F^{(j)}_t(y)\\ 
                   F^{(j)}_s(y)\end{pmatrix} \\
  &\quad\times  \bigg[\delta_{Jj} + 
    \sum_{\substack{\vec{n}\in\Zdrei\\ \vec{n}\neq\vec{0}}} \sqrt{4\pi}
    \sum_\lp i^\lp j_\lp(L|\vec{n}|y) \sqrt{\frac{(2\lj+1)(2\lp+1)}{2\lJ+1}}\\
  &\quad\times \sum_{m(\lj),s(\halb)}
    \frac{\Ct_{j,m+s}}{\Ct_{JM}} Y_{\lp (M-s-m)}^\ast(\hat{n}) 
     C^{JM}_{\lJ (M-s)\halb s} C^{j,m+s}_{\lj m\halb s}\, 
     C^{\lJ 0}_{\lj 0 \lp 0} C^{\lJ (M-s)}_{\lj m \lp (M-s-m)}
    \bigg].
\end{split}\end{equation}
\end{widetext}
The notation $m(\ell)$ is used to indicate summation over
$m=-\ell,\dots,\ell$. The partial waves of $\mathcal{Z}$ are given by
\begin{equation}
  \label{eq:Zl}
 Z^{(\ell)}(p,y) = 
 \frac{2\ell+1}{py}Q_\ell\left(\frac{p^2+y^2-E_3}{py}\right),
\end{equation}
where $Q_\ell$ is a Legendre function of the second kind. Note that
the $\delta_{Jj}$-term in Eq.~\eqref{eq:partwaves} reproduces the infinite
volume result. This equation is now specialized to the case $J=1/2$:
\begin{equation}\begin{split}
  \label{eq:jhalb}
  \begin{pmatrix}F^{(\halb)}_t(y) \\ F^{(\halb)}_s(y)\end{pmatrix} &=
    \frac{4}{\pi} \int_0^\Lambda \dd y\, y^2 \sum_j^{(G_1^+)}\left[
     \mathbf{M}_2(y)\,
      Z^{(0)}(p,y)
    + \mathbf{M}_3(y)\,
      \frac{2H(\Lambda)}{\Lambda^2}
    \right]\begin{pmatrix}F^{(j)}_t(y)\\
                          F^{(j)}_s(y)\end{pmatrix}\\
  &\quad\times\bigg[\delta_{j\halb} +
    \sum_{\substack{\vec{n}\in\Zdrei\\ \vec{n}\neq\vec{0}}} \sqrt{4\pi}
    i^{\lj} j_{\lj}(L|\vec{n}|y) \sum_{m(\lj)}(-1)^m 
    \Ct_{j,m+M} C^{j m+M}_{\lj m \halb M} 
    Y_{\lj m}(\hat{n})\bigg]\,.
\end{split}\end{equation}
Since the leading term in the expansion of the Bessel functions in
Eq.~\eqref{eq:jhalb} is $1/(L|\vec{n}|y)$, these contributions are
suppressed by at least $a/L$. They will be small for volumes not too
small compared to the size of the bound state. The lowest partial wave
that is mixed with $J=1/2$ is the $J = 7/2$ wave.
Contributions from higher partial waves will be suppressed
kinematically for shallow states with small binding momentum. This is
ensured by the spherical harmonic in the second term of
Eq.~\eqref{eq:jhalb}. Only for small lattices, i.e.  when $a/L$ is
large, this behavior is counteracted by terms stemming from the
spherical Bessel function $j_{\lj}(L|\vec{n}|y)$ and higher partial
waves may contribute significantly. In this first study, we will therefore
neglect contributions from higher partial waves. We are
encouraged to do so by our results for the bosonic case, where
calculations including one more partial wave yielded corrections on
the percent level even for small volumes~\cite{Kreuzer:2009jp}. Thus,
only two coupled integral equations remain. Details of the numerical
methods used to solve these equations can also be found
in~\cite{Kreuzer:2009jp}.

In the following, we will present our results for the energy of the
triton for cubic volumes with various edge lengths~$L$. In order to
verify that the results are properly renormalized, we produced two data sets.
For one set, the cutoff was set to $\Lambda=600$~MeV and the three-body
coupling was chosen such that the triton binding energy is reproduced
in the infinite volume. For the other set, the cutoff was chosen 
such that the three-body force vanishes for the physical triton
binding energy in the infinite volume. This is always possible due to the
limit cycle behavior of the three-body force. Both data sets should 
agree if the renormalitzation has been carried out properly.

\begin{figure}[t]
  \centering
  \includegraphics*[width=.6\columnwidth]{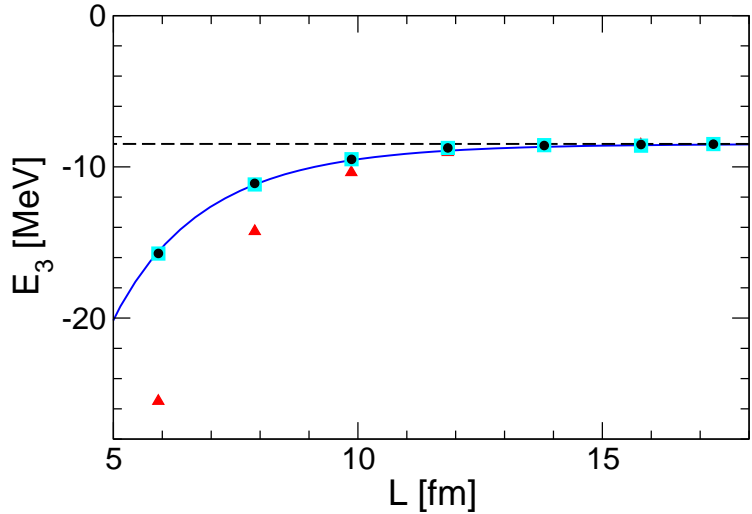}
  \caption{Triton energy $E_3$ for different edge lengths $L$ of the
    cubic volume. Squares and circles are data sets for
    different cutoffs. The triangles are from a nuclear lattice 
    calculation using chiral EFT at N$^2$LO~\cite{Epelbaum:2009zsa}.
    The horizontal dashed line indicates the physical triton binding 
    energy, while the solid line is a fit to our results (see text). 
   }\label{fig:triton}
\end{figure}
Figure~\ref{fig:triton} shows the triton binding energy for finite cubic
volumes with edge lengths $L$ ranging from 17~fm down to 2~fm. The values
from our two data sets (squares and circles) 
are in good agreement, indicating that our
results are indeed cutoff independent. For large volumes, the deviation of
the triton energy from its infinite volume value is small. When going
to edge lengths smaller than about 10~fm, the energy of the
state strongly decreases. Near $L=6$~fm, the shift is already more
than 100\%. The dependence of the energy on~$L$ can be nicely fitted
to a function of the form 
\begin{equation}
E_3(L)=E_3(L=\infty)\,\left[1+\frac{c}{L}\,\e^{-L/L_0}\right]\,,
\label{eq:Ldep}
\end{equation}
indicated by the solid line in Fig.~\ref{fig:triton}. 
We find the values $L_0 = 2.9$~fm and 
$c=39$~fm. The length scale $L_0$ is associated with the size of the physical 
triton wave function. Thus the $L$-dependence of the triton energy
is similar to the behavior in the two-body
sector~\cite{Beane:2003da,DeanPriv2010}. 

The triangles give the results from a lattice 
calculation using chiral EFT at N$^2$LO by Epelbaum and collaborators
~\cite{Epelbaum:2009zsa}. While they agree with our results for 
$L\gsim 10$ fm, the finite volume shifts are larger for smaller
volumes. This discrepancy could be due to the fact that
Epelbaum et al. calculate the volume dependence of the lowest
state in the $G_1^+$ representation, while we 
focus only on the $j=1/2$ contribution. For small volumes, the 
contribution of higher partial waves is expected to be more important.
The inclusion of higher partial waves in our approach 
is in principle straightforward but numerically expensive. In our previous
work for spinless bosons \cite{Kreuzer:2009jp}, we found the 
contribution of higher partial waves to be of the order a few
percent for $L$ about three times the size of the calculated state.
The contribution of higher partial waves in the triton will
be studied in detail in a future publication.

The next-to-leading order corrections to our results
are given by the effective ranges
in the $\sing$ and $\trip$ channels. There are corrections of
order $r_e/a$ and corrections of order $k r_e$ where $r_e$ is the 
effective range and $k$ is a typical momentum. 
The corrections of the first type are of order $30\%$.
They are dominated by the 
spin-triplet channel where the scattering length is about a factor three
larger than the effective range. The corrections of the second type
can be estimated from the typical momentum in the triton. Assuming that 
the binding momentum is shared between the nucleons in the triton,
these corrections are of order 40\% for large volumes
and grow as $L$ is decreased. 
Higher orders in the pionless EFT are required for more precise
extrapolations.

\begin{figure}[t]
  \centering
  \includegraphics*[width=.6\columnwidth]{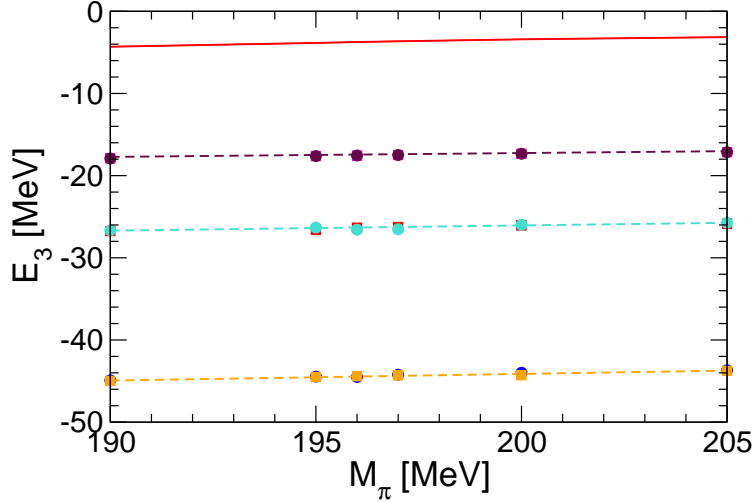}
  \caption{Pion-mass dependence of the triton energy $E_3$
   in the vicinity of the critical pion mass $\mpic\approx 197$~MeV.
   The solid line gives the infinite volume limit, while the 
   upper, middle, and lower dashed line corresponds to edge
   lengths $L=4.925$~fm, 3.94~fm, and 2.955~fm, respectively.}
\label{fig:limcyc}
\end{figure}
%
In the following, we present results for the triton and its excited
states near the critical pion mass $\mpic\approx 197$~MeV. The pion-mass
dependence of the input quantities is obtained from a chiral EFT
calculation~\cite{Epelbaum:2006jc}. These input quantities are now
used to determine the coupling constants of the ``pionless'' EFT, as
has been described for the infinite volume case
in~\cite{Hammer:2007kq}. As already stated above, no additional input is
needed to produce renormalized finite volume results. The pion mass
dependence of the triton ground state $E_3$ is shown in
Fig.~\ref{fig:limcyc} for volumes with edge lengths $L=4.925$~fm, 
3.94~fm, and 2.955~fm by the upper, middle, and lower dashed lines,
respectively. The infinite volume limit is given by the solid line.
Again, we plot two data sets obtained using two different 
cutoffs as described above which agree well with each other.
The most pronounced
effect of the finite volume is a strong downward shift, as was
expected from the results for the triton at the physical point.
Moreover, the slope of the pion-mass dependence of $E_3$ depends 
on the size of the box $L$.
From a linear fit to the pion-mass dependence 
of the triton energy in finite and infinite volume near the 
critical pion mass, we obtain the slopes
$\Delta E_3/\Delta M_\pi=$ 0.0783, 0.0467, 0.064, and 0.08
for the edge lengths $L=$ $\infty$, 4.925~fm, 3.94~fm, and 2.955~fm,
respectively.  As the volume is decreased, the slope decreases
and is smallest around $L=5$ fm. When $L$ is decreased even further
the slope increases again.

%
\begin{figure*}[t]
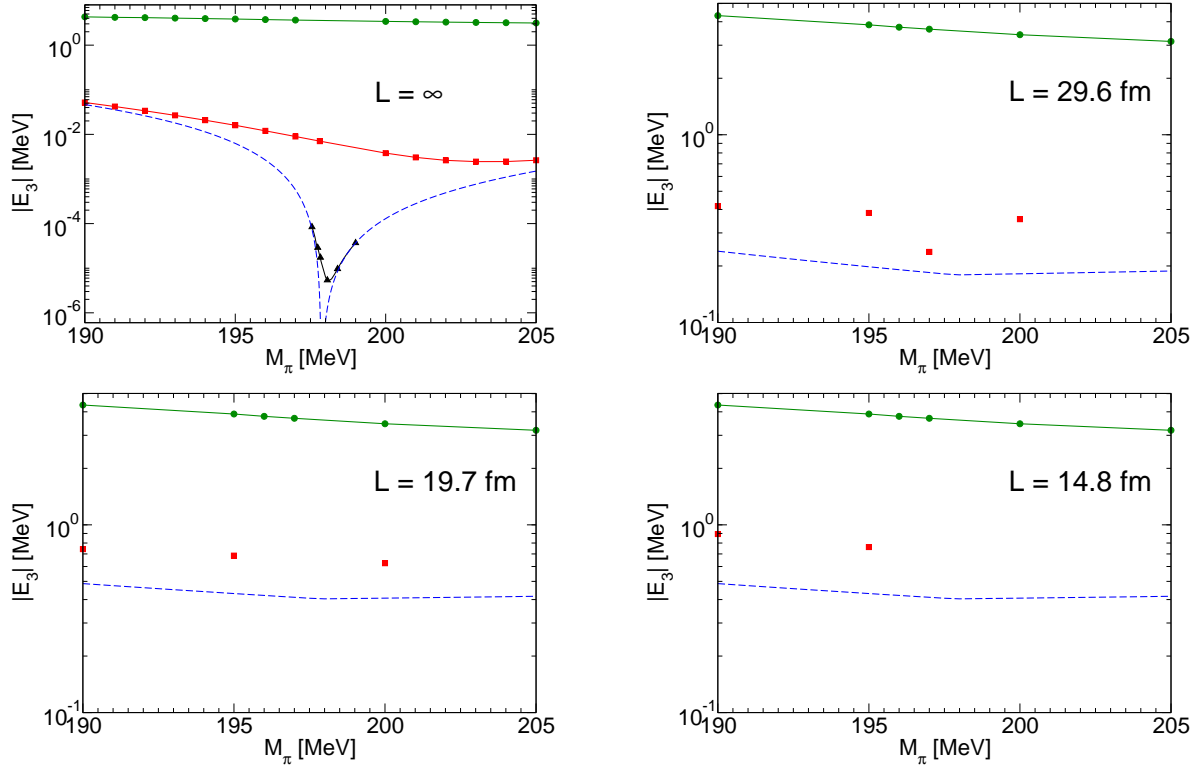

  \centering
  \qquad
  \includegraphics*[width=.44\linewidth]{limcyc_inf.eps}
  \hfill
  \includegraphics*[width=.44\linewidth]{limcyc_0.15.eps}
  \\[0.2cm]
  \qquad
  \includegraphics*[width=.44\linewidth]{limcyc_0.1.eps}
  \hfill
  \includegraphics*[width=.44\linewidth]{limcyc_0.075.eps}
  \caption{Pion-mass dependence of the triton spectrum 
    near the critical pion mass for the
    infinite volume and boxes with edge lengths 
    $L=29.6,19.7,14.8$~fm (from left to right and top to bottom). 
    The dashed lines
    are the energies of the deuteron ($M_\pi < \mpic$)
    and spin-singlet deuteron  ($M_\pi > \mpic$).
    The circles, squares, and triangles give the energies of the 
    triton ground state, first excited state, and second excited state,
    respectively.
   }
  \label{fig:spec}
\end{figure*}
The behavior of the first excited state when put inside a finite
volume is different. For large volumes, the state remains again
unaffected at first. When going to smaller volumes, the energy starts
to decrease strongly. But eventually the ``breakup threshold'', i.e.
the energy of the relevant two-body bound state in finite volume, becomes equal
to the three-body energy. For pion masses smaller than $\mpic$, the relevant
two-body bound state corresponds to the deuteron, while for
$M_\pi>\mpic$, there is a \lq\lq spin-singlet deuteron'' in the $\sing$ 
channel and the deuteron has become a virtual state. 
For even smaller volumes, no triton excited state can be found. 
This behavior is shown in Fig.~\ref{fig:spec}. 
The dashed lines give the energies of the deuteron ($M_\pi < \mpic$)
and spin-singlet deuteron  ($M_\pi > \mpic$) calculated according
to~\cite{Beane:2003da}.  The circles, squares, and triangles give the 
energies of the triton ground state, first excited state, and second 
excited state, respectively. Presumably, the triton excited 
state has crossed  into the
scattering regime where it would be driven away from threshold. We
previously observed such a behavior in our investigation of
three-boson bound states inside finite volumes~\cite{Kreuzer:2009jp}.
Whether there is a universal relation between the binding energy in
infinite volume and the volume size, at which the state disappears, is
an interesting question that will be investigated in the future. The
very weakly bound second excited state, that appears in the infinite
volume for pion masses close to the critical one, could not be
observed for the volumes with edge lengths of several fm that we
investigated and, by the above reasoning, has crossed the threshold
already at some much larger volume. 
This behavior of the excited states makes it difficult to study 
the conjectured limit cycle in Lattice QCD calculations since very
large volumes are required to observe the excited states.

In this letter, we have studied the triton inside a finite cubic
volume of edge length $L$
with periodic boundary conditions. The knowledge of the
modifications of the triton energy by a finite volume is crucial to
the understanding of lattice simulations in this channel. Using the
framework of the pionless EFT at leading order, 
we have derived an infinite set of coupled integral
equations for the partial waves of the two relevant amplitudes. This
equation was solved for several finite volumes of an order of
magnitude typical for present day Lattice QCD calculations. Proper
renormalization of the solutions was verified explicitly. 
We have investigated the
physical triton as well as the triton and its excited states near the
critical pion mass~$\mpic$, where the two-body scattering lengths in
both channels could be tuned to diverge simultaneously 
\cite{Braaten:2003eu}. We found that with present 
day volumes, the predicted excited states of the triton cannot be observed 
as bound states. 

The next step in our study is the
inclusion of higher partial waves, which is straightforward but
numerically tedious. From our previous
work for spinless bosons \cite{Kreuzer:2009jp}, we expect the
contribution of higher partial waves to be a few percent for 
$L\gsim 3 a$. Corrections from higher orders of the pionless
EFT should be included for practical applications of the volume
dependence in extrapolations. Work in these directions is in progress. 
So far, our calculations have been carried 
out for negative energies. While our theory is completely general,
the numerical treatment of positive energies above the deuteron breakup
threshold requires further work.
Finally, an estimate of finite temperature effects
would also be useful as lattice simulations are always performed at a 
small, non-zero temperature. 

We note that an 
extension of the L\"uscher formula relating the 
infinite volume scattering phase 
shifts to the discrete energy levels in a finite volume \cite{Luscher:1990ux} 
is implicitly contained in our work. It provides the framework to determine
the low-energy constants of the pionless EFT in the two- and three-body 
sector from discrete energy levels in a cubic box. After this has been
done, the infinite volume scattering observables can be calculated in 
the pionless EFT. An alternative is given by using a harmonic confinement 
instead of a cubic box. Luu and collaborators have shown how to
reconstruct the two-body scattering phase shifts from the discrete 
energy levels in harmonic confinement \cite{Luu:2010hw}. 
This method might be useful for calculations of
nuclear scattering amplitudes in effective approaches with nucleon degrees
of freedom. 

In summary, our results demonstrate that the finite volume corrections for
lattice simulations of the triton are calculable and under control.
With high statistics Lattice QCD simulations of three-baryon systems
within reach~\cite{Beane:2009gs}, the calculation of the structure and
reactions of light nuclei appears feasible in the intermediate
future. 

\begin{acknowledgments}
We thank Dean Lee, Ulf Mei\ss ner, Daniel Phillips, and Martin Savage
for discussions. We acknowledge the hospitality of the 
Institute for Nuclear Theory in Seattle,
where part of this work was carried out.
This research was supported by the DFG through
SFB/TR 16 \lq\lq Subnuclear structure of matter'' and the BMBF
under contracts No. 06BN9006.
\end{acknowledgments}


\end{document}